\documentclass[letterpaper,twocolumn,10pt]{article}
\usepackage{usenix}

\usepackage{tikz}
\usepackage{amsmath,amssymb,amsfonts}
\usepackage{algorithmic}
\usepackage{graphicx}
\usepackage{textcomp}
\usepackage{xcolor}
\usepackage{listings}
\usepackage{caption}
\usepackage{subcaption}
\usepackage{authblk}
\usepackage{afterpage}
\usepackage{float}

\definecolor{codegreen}{rgb}{0,0.6,0}
\definecolor{codegray}{rgb}{0.5,0.5,0.5}
\definecolor{codepurple}{rgb}{0.58,0,0.82}
\definecolor{backcolour}{rgb}{0.95,0.95,0.92}

\lstdefinestyle{mystyle}{
    backgroundcolor=\color{backcolour},
    commentstyle=\color{codegreen},
    keywordstyle=\color{magenta},
    numberstyle=\tiny\color{codegray},
    stringstyle=\color{codepurple},
    basicstyle=\ttfamily\footnotesize,
    breakatwhitespace=false,
    breaklines=true,
    captionpos=b,
    keepspaces=true,
    numbers=left,
    numbersep=5pt,
    showspaces=false,
    showstringspaces=false,
    showtabs=false,
    tabsize=2
}

\begin{document}

\date{}

\title{\Large \bf Data Caching for Enterprise-Grade Petabyte-Scale OLAP}

\makeatletter
\renewcommand\AB@affilsepx{, \protect\Affilfont}
\makeatother

\author[1]{Chunxu Tang}
\author[1]{Bin Fan}
\author[2]{Jing Zhao}
\author[2]{Chen Liang}
\author[1]{Yi Wang}
\author[1]{Beinan Wang}
\author[3,2]{Ziyue Qiu}
\author[1]{Lu Qiu}
\author[1]{Bowen Ding}
\author[1]{Shouzhuo Sun}
\author[1]{Saiguang Che}
\author[1]{Jiaming Mai}
\author[1]{Shouwei Chen}
\author[1]{Yu Zhu}
\author[1]{Jianjian Xie}
\author[4]{Yutian (James) Sun}
\author[2]{Yao Li}
\author[2]{Yangjun Zhang}
\author[4]{Ke Wang}
\author[2]{Mingmin Chen}

\affil[1]{Alluxio, Inc.}
\affil[2]{Uber, Inc.}
\affil[3]{Carnegie Mellon University}
\affil[4]{Meta, Inc.}

\maketitle

\begin{abstract}

With the exponential growth of data and evolving use cases, petabyte-scale OLAP data platforms are increasingly adopting a model that decouples compute from storage. This shift, evident in organizations like Uber and Meta, introduces operational challenges including massive, read-heavy I/O traffic with potential throttling, as well as skewed and fragmented data access patterns. Addressing these challenges, this paper introduces the Alluxio local (edge) cache, a highly effective architectural optimization tailored for such environments. This embeddable cache, optimized for petabyte-scale data analytics, leverages local SSD resources to alleviate network I/O and API call pressures, significantly improving data transfer efficiency. Integrated with OLAP systems like Presto and storage services like HDFS, the Alluxio local cache has demonstrated its effectiveness in handling large-scale, enterprise-grade workloads over three years of deployment at Uber and Meta. We share insights and operational experiences in implementing these optimizations, providing valuable perspectives on managing modern, massive-scale OLAP workloads.

\end{abstract}
\section{Introduction}

With the exponential growth of data and the continuous evolution of use cases, modern online analytical processing (OLAP) systems have become crucial in many data-centric organizations, such as Uber, Meta, and Twitter. Historically, many interactive OLAP engines were designed with co-located compute and storage, necessitating the pre-loading of data into local SSDs, HDDs, or even memory to enhance querying performance. While this approach was effective in certain contexts, it led to significant challenges, including storage fragmentation, data duplication, and stranded resources. In response, there has been a fundamental shift to a more versatile architecture that decouples compute and data lake storage \cite{tan2019choosing, chattopadhyay2023shared}. With this disaggregation, data, metadata, and compute are allowed to scale independently, thereby enhancing efficiency and flexibility in handling diverse and large-scale workloads.

\begin{figure}[t]
    \centerline{\includegraphics[width=0.47\textwidth]{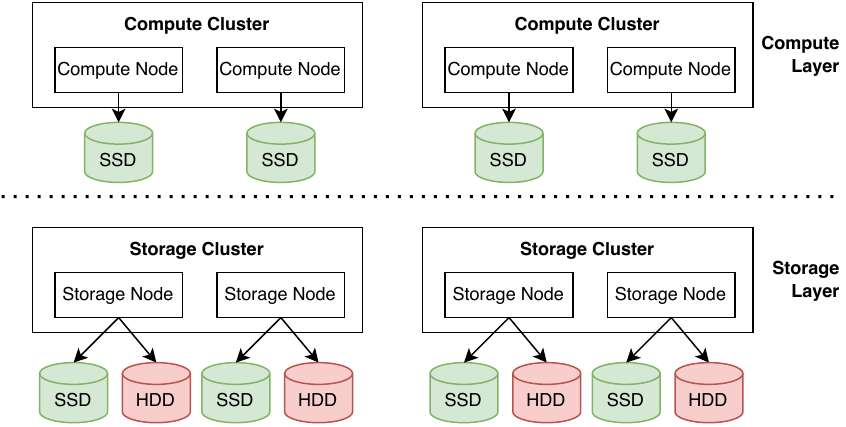}}
    \caption{Compute-storage disaggregation with SSD cache.}
    \label{fig.high-level}
\end{figure}

This architectural evolution towards decoupled compute and data lake storage, however, introduces new operational challenges for large-scale OLAP systems like Presto \cite{sethi2019presto}, Redshift \cite{gupta2015amazon}, and BigQuery \cite{bigquery}. A primary challenge lies in the significant strain on network resources or API throughput of the data lake, driven by the vast volumes of data accessed by these systems. Additionally, while the total amount of data is rapidly expanding, we observe pronounced skewness in access: a few hot files or data blocks account for the majority of the traffic. Moreover, the prevalence of columnar data formats in modern data lakes, while beneficial for interactive queries, leads to the need for small, disparate reads instead of sequentially reading large chunks of data. Our study of data analytics and storage traces at companies like Uber has identified these intertwined challenges – disproportionate read traffic, data access hot spots, and network resource strain – all contributing to the complex landscape of managing data in these evolved systems.

While the challenges are complex and multifaceted, introducing major software or hardware changes at the platform level, or fundamentally redesigning applications, query engines, or storage engines, can be a lengthy process. Such changes often take years and may significantly impact critical business operations, sometimes proving to be nearly impossible. Consequently, with joint efforts from Alluxio, Uber, and Meta, we experimented and implemented incremental and isolated I/O optimizations within the architecture. In particular, we introduce the \textit{Alluxio local (edge) cache}, an embeddable cache suitable for both compute and storage engines in a non-disruptive manner (Figure \ref{fig.high-level}). Designed to leverage local SSD resources, the Alluxio local cache not only effectively minimizes slow network or hard-disk data retrieval associated with repetitive I/O requests but also supports both random and sequential data access in OLAP \cite{sirin2016micro}, thanks to its SSD-based, page-oriented structure optimized for high throughput and concurrency in petabyte-scale OLAP.

Being a collaboration of open-source communities across different organizations, this project intentionally kept design choices simple and incremental to facilitate rapid iteration and quick consensus among multiple parties. Surprisingly, our relatively straightforward designs and optimization significantly impact business-critical, enterprise-grade platforms at a petabyte scale. We observed that adding such a local cache to both compute and storage effectively leverages the strong temporal and spatial locality of modern OLAP workloads, significantly reducing request latency and improving data transfer efficiency between compute and storage layers. After over three years of deployment at Uber \cite{uber-presto,uber-hdfs} and Meta \cite{meta-alluxio}, Alluxio local cache has proven effective in handling Internet-scale, enterprise-grade workloads. Specifically, the integration of Alluxio local cache within Presto, known as \textit{Presto local cache}, has decreased the 95th percentile (P95) of query latency by 49\% in Meta's production data analytics. Similarly, the implementation of Alluxio local cache in HDFS, referred to as \textit{HDFS local cache}, has reduced the number of processes blocked due to I/O throttling by 86\% in Uber's production HDFS clusters.

Our contributions in this paper include:
\begin{itemize}
    \item We provide an analysis of real-world trace data collected from Uber's OLAP and storage platforms, focusing on I/O and their implications.
    \item We detail the requirements, design philosophy, and implementation of the Alluxio local cache, illustrating its role as an effective optimization for petabyte-scale OLAP and storage systems.
    \item We provide valuable insights stemming from our firsthand industrial experience in supporting and maintaining this optimized architecture over several years.
\end{itemize}

The remainder of this paper is organized as follows: Section \ref{sec.background} explores industrial data analytical and storage systems, outlining key workload characteristics derived from production traces that motivate the Alluxio local cache design. Section \ref{sec.requirements} details practical design requirements for the cache system. Section \ref{sec.arch} elaborates on the architectural design and primary components of Alluxio local cache. Optimizations and enhancements are discussed in Section \ref{sec.optimization}. Section \ref{sec.use-case} presents case studies on the application of Alluxio local cache in Presto and HDFS. Sections \ref{sec.lessons} and \ref{sec.failures} reflect on lessons learned and failure case studies from various developmental and operational contexts. Section \ref{sec.related-work} reviews related work in this domain, and Section \ref{sec.conclusion} concludes the paper.

\section{Background}\label{sec.background}

This section provides an in-depth examination of the complexities and challenges associated with large-scale data analytics in an enterprise environment, enumerating Uber as a primary case study. It offers a detailed exploration of enterprise-grade data analytical and storage systems, emphasizing the specific characteristics and challenges posed by managing extensive datasets. Given the data privacy regulations adhered to by various organizations, our analysis primarily focuses on the real-world workloads observed in Uber's data analytical and storage operations. Additionally, to provide a broader perspective and supplementary insights, the workloads and data handling strategies of other enterprises, such as Meta and Twitter, are also considered.

\subsection{Uber's Data Analytical \& Storage Systems}

As a prime example of a modern data-centric corporation, Uber's reliance on analytics to inform its vast array of services cannot be overstated. Enumerating Uber's data infrastructure as an example, we elaborate on typical enterprise-grade data analytical and storage systems.

\subsubsection{Data Query}

For Uber, data isn't just instrumental--it's the backbone of decision-making. At the heart of this data-driven model is Presto. Presto is an open-source distributed SQL query engine for OLAP. It employs a coordinator-worker architecture: a central coordinator node takes charge of parsing queries, formulating query plans, and distributing tasks to worker nodes; and multiple worker nodes (can be as high as thousands) handle data retrieval from storage sources, such as on-premises HDFS or cloud-based storage like GCS.

Presto has been adopted for a wide spectrum of use cases. At Uber, whether it is the operations team synthesizing real-time dashboards or the Uber Eats and marketing divisions making pricing calls, Presto's influence permeates. Units like compliance and growth marketing also rely on insights gathered from Presto SQL queries. Besides Uber, Presto is embraced by a large number of companies: Meta uses Presto for diverse applications, including dashboard reporting, ad-hoc analysis, and data transformation. Meta engineers also combined Presto and Spark within the Presto on Spark project to use Presto as the single piece to serve interactive, ad-hoc, extract-transform-load (ETL), and graph processing jobs \cite{sun2023presto}. Presto dominated Twitter's interactive analytics for hate speech analysis, engagement data validation, and tool dashboard reporting \cite{tang2022serving}.

At Uber, the sheer magnitude of Presto's operation is evident: approximately 9,000 daily users, around 500,000 daily queries, and a management of 50PB of data. This expansive setup spans 2 data centers, 7,000 nodes, and 20 unique Presto clusters across two regions \cite{presto-uber}.

\subsubsection{Data Storage}

With one of the most formidable Hadoop Distributed File System (HDFS) deployments globally, Uber's exabyte-spanning data across various clusters demands persistent scalability. Achieving a harmony between efficiency, reliability, and performance is critical and challenging. HDFS powers a wide range of use cases at Uber, including fraud detection, machine learning, and expected time of arrival (ETA) calculation. The data analytical requests dominate HDFS traffic, which is one of the motivations for implementing a cache system for both Uber's data analytical and storage systems.

To scale Uber's HDFS, Uber engineers instituted several enhancements, such as the adoption of View File System (ViewFs), NameNode garbage collection tuning, and an HDFS load management service \cite{uber-hdfs-old}. For further cost efficiency, a transition to denser HDDs (16+TB SKUs) is underway, superseding the current 4TB HDD SKUs that dominate the HDFS clusters. This indicates that HDD capacities will expand by factors of 2x to 4x \cite{uber-hdfs}.

\subsection{Industrial Data Analytical \& Storage Workload Characteristics \& Challenges}\label{sec.background.workload}

With a closer look at enterprise-grade workloads for interactive query processing and large-scale data storage, we observed a few crucial workload characteristics and challenges, which significantly motivated the work of Alluxio local cache.

\paragraph{Data-intensive Analytics at Massive-Scale}

Our analysis highlights the massive scale of traffic in enterprise-grade data analytics and storage workloads. For instance, Uber's Presto system processes an astonishing volume of data, handling around 500 thousand queries and managing 50 petabytes (PB) of data daily \cite{presto-uber}. Similarly, Twitter's hybrid-cloud SQL federation system addresses tens of thousands of queries each day, involving roughly 10 PB of data \cite{tang2021forecasting}. Additionally, Meta’s extensive deployment of Presto across thousands of nodes processes hundreds of PBs of data on a daily basis \cite{luo2022batch}.

Table \ref{tab.uber-hdfs} presents statistics from four HDFS DataNodes in high-activity production clusters, reflecting the scale of data transactions over a period of approximately 20 hours. These statistics underscore the substantial scale of operations and the need for advanced solutions to manage such large-scale data workloads. Notably, at Uber, HDFS traffic is primarily driven by OLAP queries, with Presto queries alone accounting for approximately 90\% of this traffic.

\begin{table}[htb]
    \begin{tabular}{|c|c|c|c|c|}
    \hline
    Host                                                                      & Host1  & Host 2 & Host 3 & Host 4 \\ \hline
    Total reads (M)                                                           & 13.5   & 12.8   & 8.5    & 14.3   \\ \hline
    Total writes (K)                                                          & 3.3    & 4.7    & 4.6    & 45     \\ \hline
    Reads / writes                                                            & 4091.0 & 2723.4 & 1847.8 & 317.8  \\ \hline
    \begin{tabular}[c]{@{}c@{}}Read traffic on \\ top 10K blocks\end{tabular} & 89\%   & 94\%   & 99\%   & 99\%   \\ \hline
    \end{tabular}
    \caption{Production traffic of Uber's HDFS clusters.}
    \label{tab.uber-hdfs}
\end{table}

\paragraph{Skewness in Data Access Patterns}

\begin{figure}[thp]
    \centerline{\includegraphics[width=0.24\textwidth]{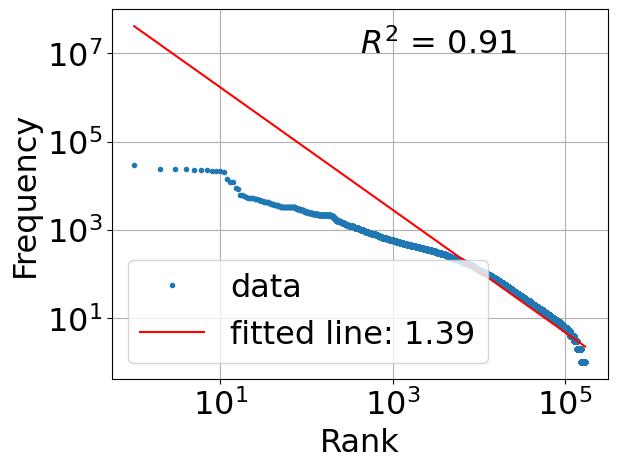}}
    \caption{Popularity rank and Zipfian distribution, indicating a heavy skewness in data access distribution.}
    \label{fig.skewness}
\end{figure}

A deeper analysis of enterprise-grade workloads reveals a pronounced skewness in read data access distribution. For illustration, Figure \ref{fig.skewness} demonstrates that an average Presto node at Uber has a Zipfian factor of up to 1.39, indicating heavily skewed data access patterns. This means that some files are accessed with disproportionately high frequency, highlighting the potential efficiency gains achievable through caching.

Similarly, in Uber's HDFS traffic, distinct hot spots are observed. Table \ref{tab.uber-hdfs} provides a summary of the read traffic for the top 10K data blocks across four HDFS DataNodes. This data indicates that up to 99\% of the read traffic is concentrated on these top 10K blocks, further emphasizing the importance of effective data management strategies.

\paragraph{Limited I/O Bandwidth with High-Density Disks}

Adopting high-density disk SKUs presents challenges, particularly in terms of disk I/O bandwidth. For instance, while Uber's HDD capacities are expanding by 2x to 4x, the I/O bandwidth of each HDD isn't increasing at a comparable rate. This discrepancy results in I/O throttling within HDFS DataNodes, which is evident from the number of blocked processes on a node. In Uber's production traffic, the count of blocked processes can reach up to several thousand within just one minute. Given the persistent and significant instances of slow reads, preventing performance degradation becomes crucial.

\paragraph{Fragmented Access to Columnar Files}

In modern data lakes, the data processed by distributed SQL engines like Presto and Spark SQL is increasingly converging to columnar formats such as ORC \cite{apache-orc} or Parquet \cite{vohra2016apache}. These formats are favored for their efficiency in read operations. These engines have implemented various query optimization techniques, with predicate pushdown being a key example. This technique allows the system to move filtering conditions closer to the storage layer. While these optimizations lead to performance gains, they also often result in a high number of read requests for small portions of data files.

This shift is clearly reflected in Uber's HDFS production traces. More than 50\% of SQL requests on HDFS access less than 10KB of data, and over 90\% involve less than 1 MB \cite{tang2023rethinking}. Such statistics underscore the importance of page-based cache optimizations and strategic management in fragmented data access, especially for small data chunks.

\section{Design Requirements}\label{sec.requirements}

In addition to the aforementioned workload characteristics and challenges that motivate the design and implementation of a cache system, there are a few other design requirements listed as follows.

\textbf{An application-agnostic cache.} The prevalence of the aforementioned characteristics across various data analytical and storage workloads necessitates the development of a cache system that is both flexible and extensible. Such a system should be capable of integrating with a range of applications and use cases, thus requiring an application-agnostic approach. We present the architectural design of Alluxio local cache in Section \ref{sec.arch}.

\textbf{Minimal socket communication with applications.} Fetching data over the network is slow for interactive analytics scenarios \cite{chattopadhyay2023shared}. The cache system needs to minimize socket communication with the application leveraging the cache. Due to this requirement and resource limitations, we preferred a local cache system over a distributed cache. The local cache runs within the same JVM process as the application, interacting with the remote storage through a Hadoop-compatible interface. Further details are discussed in Section \ref{sec.arch.overall}.

\textbf{Non-volatile data storage.} Given the substantial memory usage by modern OLAP systems, primarily for in-memory storage of intermediate results, the cache system is expected to have a minimal memory footprint. By design, we aim for single-node caching of up to 1 TB of data. The high volume of data processed, combined with memory constraints, renders in-memory caching of all data files impractical. We shed light on the storage medium for cache data storage in Section \ref{sec.arch.storage}.

\textbf{Capability to support both random access and sequential data access.} Industrial data analytical and storage workloads are inclined to frequently access small data segments. Therefore, the cache system must efficiently support not only traditional sequential data access but also random access, which targets specific portions of a data file. A page-based organization of cached data is an effective way to meet this need. We present details in Sections \ref{sec.arch.page} and \ref{sec.arch.indexedset}.

\textbf{High performance.} Traditionally, a cache system is expected to be highly performant, aiding in reducing the request latency through the storage of frequently accessed data in a quicker-access medium. This conventional cache design requirement remains essential. We discuss details on various cache optimizations and enhancements in Section \ref{sec.optimization}.

Besides the above requirements, when an enterprise-grade large-scale data system adopts a local cache, the local cache will inevitably face classical distributed systems' challenges, encompassing aspects like reliability, scalability, and availability. These challenges foster adaptations and further development of the local cache, such as soft-affinity scheduling, application-level metrics aggregation, and cache rate limiting strategies. We elaborate on concrete practical challenges and corresponding implementation solutions in Section \ref{sec.use-case}.

\section{Architectural Design}\label{sec.arch}

\subsection{Overall Design}\label{sec.arch.overall}

\begin{figure}[thp]
  \centerline{\includegraphics[width=0.34\textwidth]{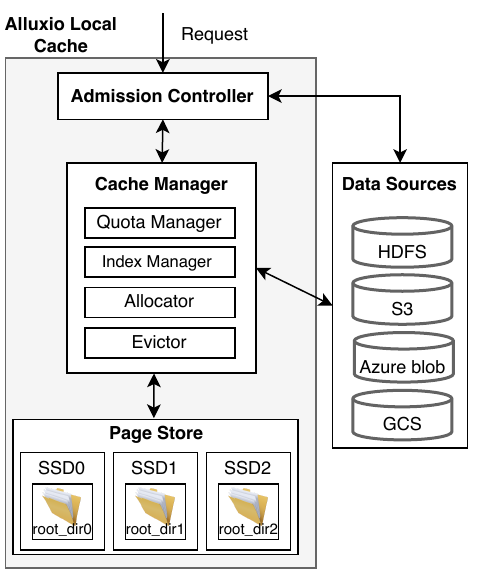}}
  \caption{The workflow of Alluxio local cache.}
  \label{fig.overall-arch}
\end{figure}

Figure \ref{fig.overall-arch} depicts the comprehensive workflow of handling data requests in Alluxio local cache, which extends the file system APIs from Alluxio \cite{li2018alluxio,li2014tachyon}. The local cache is primarily designed to enhance read performance by introducing a straightforward yet effective local cache layer. The workflow involves an \textit{admission controller}, a pivotal component that identifies and prioritizes the most frequently accessed files for caching. The controller ensures that only data deemed essential for caching is admitted, while other data leads to a non-cache read path, directed to external \textit{data sources}.

Alluxio local cache transforms file-level read operations into more granular page-level operations through the \textit{page store}. The page store comprises multiple local cache directories, each with its designated cache capacity. Local SSDs are the recommended storage medium for cached data, considering their performance efficiency.

Central to the Alluxio local cache is the \textit{cache manager}, which encompasses the system's key functionalities. When a requested page is found in the cache (cache hit), the cache manager promptly returns the requested page to the client. Conversely, in cases of cache misses, the page is fetched from external data sources, cached locally, and served to the client (read-through strategy). The \textit{quota manager} maintains a hierarchical resource control management, allowing for quota specifications at various levels, such as global, schema, table, and partition levels. The \textit{index manager} organizes pages efficiently based on page metadata criteria, thereby optimizing the conditional page search to O(1) complexity. The \textit{allocator} is responsible for assigning cache pages to appropriate directories, considering factors like file identification, hash algorithms, directory capacity, and page affinity. In scenarios where a directory's available space is insufficient or if a quota rule is breached, the system initiates an eviction process. The \textit{evictor} component orchestrates multiple cache eviction strategies, such as FIFO, random, and LRU. It provides an interface for the integration of alternative policies if needed. Besides the conventional policy-based cache eviction, the local cache also incorporates a time-based eviction strategy by setting the time to live for each cached file. A periodic background job evicts expired data as needed. This feature is raised and adopted in response to the growing data privacy requirements.

\subsection{Cache Data Storage}\label{sec.arch.storage}

Currently, many cache systems prefer to store data in memory. The high speed of memory access, especially random data access, makes this design suitable for extremely latency-sensitive production scenarios. For example, Twitter developed Twemcache based on the open-sourced Memcached, which serves as an in-memory key-value cache to serve real-time user and tweet data \cite{yang2021large}. Similarly, Meta scaled and enhanced Memcached to manage billions of requests per second in vast social media networks \cite{nishtala2013scaling}.

Nonetheless, the storage medium of the local cache is constrained by the available memory in the node. Modern enterprise-grade large-scale OLAP systems are known for intensive memory usage, leaving limited room for additional cache storage within the memory. Although scaling up the local node's hardware resources is a viable solution to augment higher memory capacity for caching, this solution can incur substantially higher costs, particularly when scaling to serve large volumes of enterprise-grade data analytics.

SSDs emerged as an alternative solution for caching data locally. SSDs present an advantageous balance between cost-efficiency and high-speed access for both random and sequential data access. An analysis of resource utilization in data analytical jobs at Uber and Meta reveals that local SSDs are usually underutilized, indicating a potential opportunity for cache optimization. Consequently, these insights guide our decision to leverage local SSDs for caching data, while maintaining the metadata still in memory to ensure fast access.

\subsection{Page-Based Organization}\label{sec.arch.page}

Alluxio local cache organizes cached data in pages. Initially, the default page size was 64 MB, the same as the default block size in HDFS. However, from our operational experience on effective support of columnar files and SSD bandwidth, we adjusted the default page size to 1 MB (Section \ref{sec.lessons}). In practice, a production data file, such as a Parquet file at Uber, is cached as hundreds of pages, providing finer-grained control on portions of the file. When the write operation of a page is completed, the cached page will be immediately available for subsequent read operations. We developed fine-grained locking mechanisms to support high-read concurrency. Page information is self-contained in page names and parent folders, so the application can identify whether the data it wants is cached by simply looking up the metadata information.

\begin{figure}[thp]
    \centerline{\includegraphics[width=0.33\textwidth]{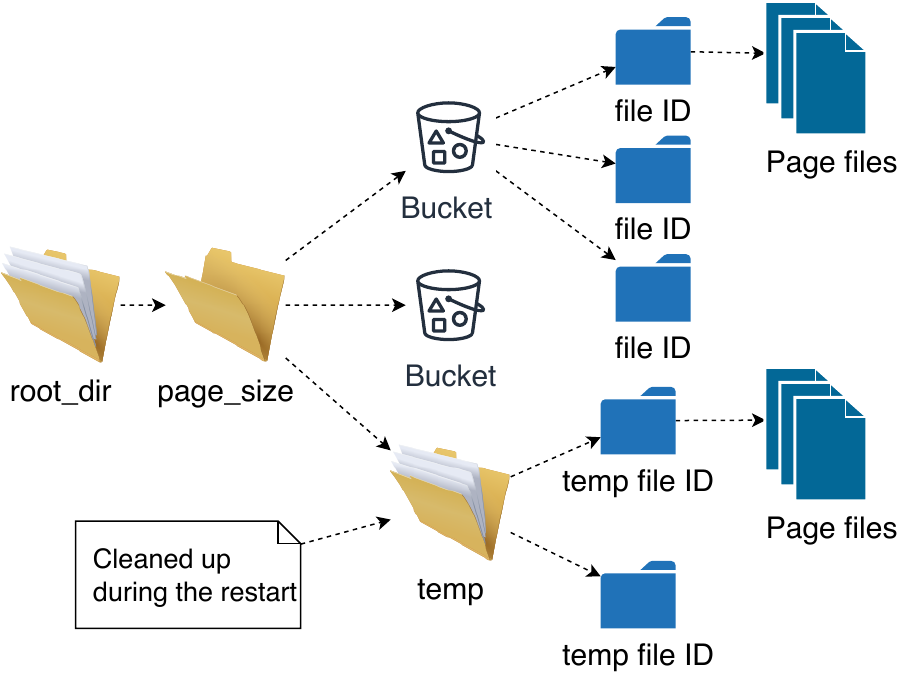}}
    \caption{Organization of cached data files.}
    \label{fig.dir-structure}
\end{figure}

Additionally, cached data is organized in a multi-level hierarchy, as shown in Figure \ref{fig.dir-structure}. Shared information is organized as folders. Top-level folders represent persistent global information that can be used in cache recovery. For example, the \textit{page\_size} folder is a top-level folder, since the page size information is required to calculate the page index, which is subsequently used together with file IDs to compute page IDs. Shared file information is stored as folders, such as full paths, and file version information. To avoid one directory containing too many sub-folders or files to downgrade the lookup performance, we 1) add buckets for an additional layer and 2) add file ID directories.

\subsection{Index Management}\label{sec.arch.indexedset}

Alluxio local cache supports common cache management operations such as read, write, and delete on different granularity levels. The basic levels include the page level, being the finest level where each page can be addressed individually, and the file level composed of the pages belonging to the same file.

Logical levels can be leveraged by applications to perform bulk page operations efficiently. For example, in Presto, the data is organized in a partition-table-schema hierarchy. This hierarchy maps to a tree of nested scopes in Alluxio local cache: a global scope representing the entirety of the cache and scopes for each schema, table, or partition. Pages belonging to a particular data file are assigned with scopes of the partition, table, and schema of that file. Users can perform I/O operations on all pages of a specified scope efficiently. For example, in a table partitioned by a date column, a user may wish to delete all pages belonging to a certain outdated partition, to free up space for more up-to-date data. Without the partition scopes, the user would otherwise have to first find all files that belong to that specific partition, possibly by invoking a costly directory listing operation and filtering through the result to locate all the files.

\begin{figure}[thp]
  \centerline{\includegraphics[width=0.45\textwidth]{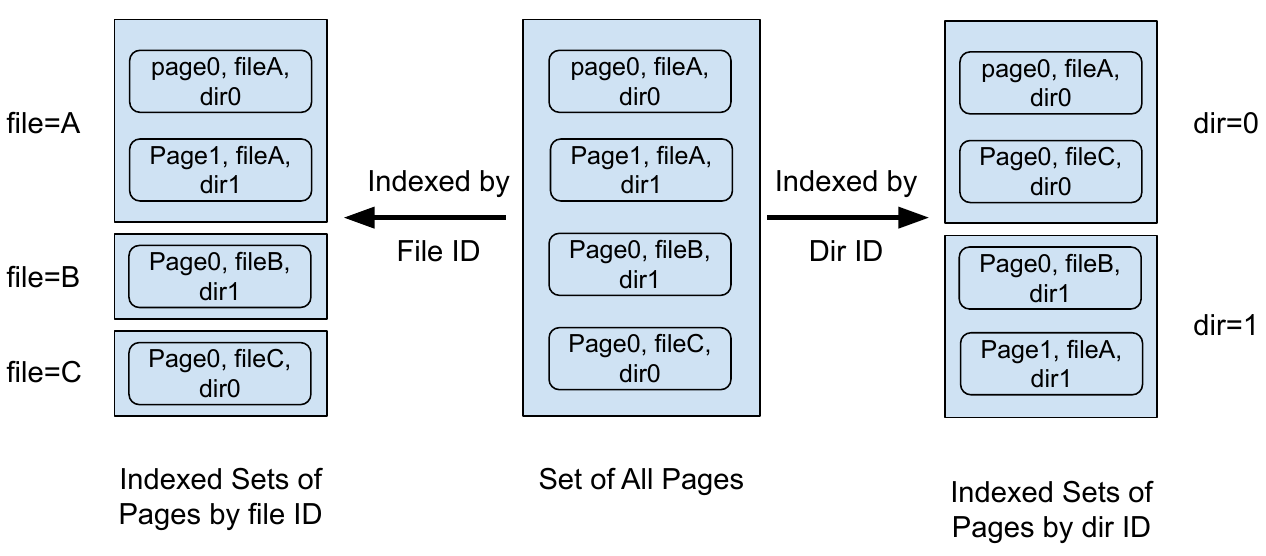}}
  \caption{Using indexed sets for organizing pages.}
  \label{fig.indexed-set}
\end{figure}

There is also a level that corresponds to the medium of the page store. This level allows users to address all pages stored in a particular storage device, mounted as a directory on the file system of the host. This is typically useful for reporting the cache usage on a particular device, or deleting all pages in a storage device, when the backing device becomes faulty.

Initially, all the page metadata was stored in two maps, indexed by the page ID and the ID of the files they are part of, respectively. This naive approach quickly proved insufficient as the need to support more diverse levels became prominent. To support efficiently accessing pages for a given level, without resorting to iterating over the entire page store, the pages need to be properly indexed with the different levels. We use indexed sets to store all pages' metadata. As Figure \ref{fig.indexed-set} illustrates, the universe set contains all pages that are currently stored in the cache. Each indexed set is a subset of the universe indexed by a certain property of the page's metadata. There are three sets indexed by the file ID A, B, and C, and three sets indexed by storage directories 1 and 2, respectively.

\section{Cache Optimizations \& Improvements}\label{sec.optimization}

\subsection{Cache Admission Strategies}\label{sec.arch.filter}

Nowadays, a modern OLAP system is capable of handling petabyte-scale data daily, making storing all production datasets in the local cache unfeasible. As reported in Section \ref{sec.background.workload}, the frequency of data block access varies significantly, with certain hot data blocks accessed hundreds of thousands of times within a single day. Consequently, the admission decisions are governed by several strategies.

One such approach defines cache admission criteria as regular expressions or more versatile JSON-format expressions. This strategy has been effectively used in the Presto local cache (Section \ref{sec.use-case.presto}). For instance, the following code snippet illustrates a static filtering rule applied to the \textit{table\_bar} table. The \textit{maxCachedPartitions} field specifies the upper limit of 100 on the number of data partitions of the table that can be retained within the cache. In production, the filtering rules are set by platform owners and infrequently updated. At Uber, after such filtering, less than 10\% of requests require remote storage access \cite{presto-uber}.

\begin{lstlisting}
{
  "databases": [
    {
      "name": "database_foo",
      "tables": [
        {
          "name": "table_bar",
          "maxCachedPartitions": 100
        }
      ]
    }
  ]
}
\end{lstlisting}

An alternative solution involves the analysis of historical data access patterns in a defined time frame. This admission strategy leverages sliding windows to assess the hotness of data blocks. This approach proves its effectiveness in HDFS local cache (Section \ref{sec.use-case.hdfs}). For the requests which fulfill the admission policy, only around 1\% of them require slower storage access.

\subsection{Quota Management for Multi-Tenancy}\label{sec.arch.quota}

The support of multi-tenancy is a common need in an enterprise-grade cache system. To enable fairness and efficient resource utilization, Alluxio local cache adopts flexible multi-layer quota management, preventing any single tenant from excessively consuming resources. It facilitates the allocation of quotas through a hierarchical tenant-based approach: 1) user tenants, where quotas are set for different teams, organizational units, or regions; 2) data tenants, where resource allocation is based on specific data entities like partitions or tables; 3) custom tenants, offering flexibility for bespoke quota configurations based on any logical grouping, such as project-based or application-specific.

The quota verification process in Alluxio local cache is hierarchical, starting from the most detailed level (often partitions) and ascending through tables, schemas, and up to the global level. This quota check prevents any single tenant from monopolizing cache resources and starving others.

Our initial implementation restricted the total quota for a table's partitions to not exceed the table's quota. However, practical experience in a production environment revealed that this limitation hindered efficient resource sharing. Consequently, we evolved the design to allow the collective quota of partitions to surpass the quota of their parent table. For instance, if a table with a 1 TB quota has two partitions, each with an 800 GB quota, the system ensures that no partition exceeds its individual quota (800 GB).

In situations where a quota violation occurs, Alluxio local cache evicts data to re-establish quota compliance. There are two eviction strategies: 1) Partition-level eviction. If a partition exceeds its quota, data eviction is triggered for that specific partition. 2) Table-level sharing and eviction. If the overall usage at the table level exceeds its quota, the system performs random eviction across partitions. This randomization is particularly beneficial in cases where one partition demands significantly more cache space than others, ensuring efficient resource sharing and utilization.

\section{Use Cases \& Evaluation}\label{sec.use-case}

\begin{figure}[htp]
    \centerline{\includegraphics[width=0.48\textwidth]{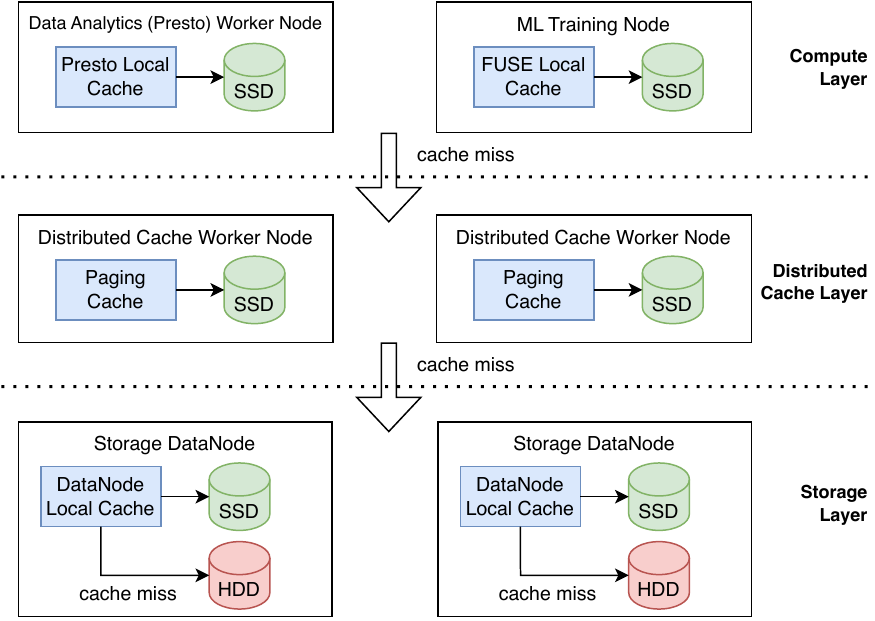}}
    \caption{Alluxio local cache use cases, which cover the compute, distributed cache, and storage layers.}
    \label{fig.arch}
\end{figure}

This section discusses practical use cases and evaluation of Alluxio local cache. Thanks to years of efforts from open-source communities, nowadays, the original Alluxio local cache has evolved into a more mature, robust system that contains advanced features, such as caching filtering, scheduling, and quota management for multi-tenancy. Alluxio local cache has been used in scenarios that are well beyond the initially designed boundaries of SQL query engines like Presto. It functions as a general-purpose local cache and finds applications across a diverse range of scenarios including distributed file systems, machine learning training, and distributed cache.

As depicted in Figure \ref{fig.arch}, at the compute layer, in the data analytics landscape, query engines like Presto use Alluxio local cache to enhance performance and cost-efficiency. Meanwhile, in the realm of machine learning, particularly in training phases, Filesystem in Userspace (FUSE) \cite{fuse} utilizes the local cache to help improve training performance and GPU utilization. At the storage layer, HDFS DataNodes leverage Alluxio local cache within each node to bolster data retrieval performance and mitigate I/O bottlenecks. Bridging the compute layer and the storage layer is a distributed cache layer, where Alluxio local cache is integrated into each cache worker node to serve the traffic. This section primarily focuses on two key aspects: the local cache applied for data analytics (Presto local cache) and the local cache used for storage (HDFS local cache). A detailed discussion on the application of local cache in machine learning training and the distributed cache layer will be presented in a following paper.

\subsection{Case Study - Presto Local Cache}\label{sec.use-case.presto}

\subsubsection{Design Overview}

The journey of integrating Alluxio local cache into Presto started in 2019. It is initially envisioned as a pluggable and lightweight cache module, aiming at decreasing query latency and reducing the data transfer between the compute and storage tiers. Contributions from open-source communities of Presto and Alluxio, alongside inputs from tech companies like Uber, Meta, and Twitter, have significantly evolved Presto local cache to fulfill the growing demand for data analytics at the enterprise-grade. Presto supports reading and parsing various file formats, particularly column-oriented file formats widely used in OLAP, such as Apache Parquet \cite{vohra2016apache} and Apache ORC \cite{apache-orc}. These column-oriented file formats inherently organize data files by columns and expedite data processing by segmenting data into row groups (stripes in ORC, row groups in Parquet). Presto employs such design and distributes splits among the worker nodes following the generation of a distributed query plan by the coordinator node.

\begin{figure}[thp]
    \centerline{\includegraphics[width=0.45\textwidth]{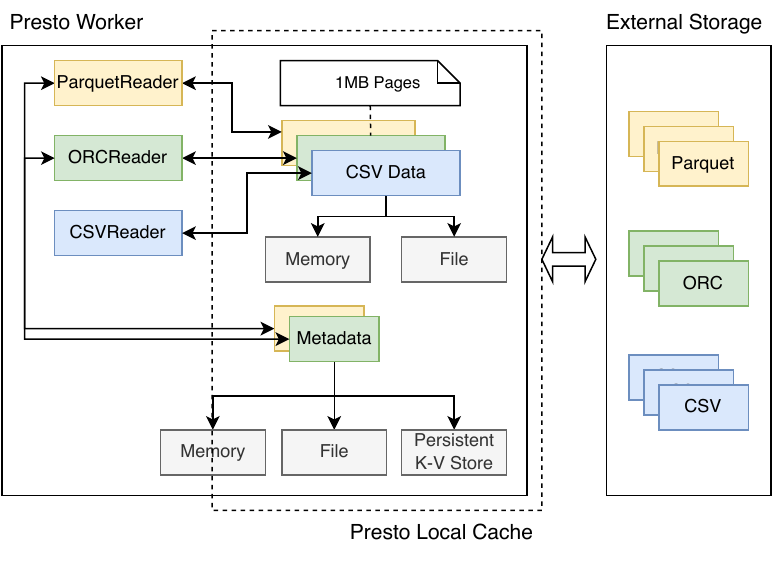}}
    \caption{Caching data and metadata in Presto local cache.}
    \label{fig.presto-sdk}
\end{figure}

As shown in Figure \ref{fig.presto-sdk}, Presto incorporates specific reader classes to handle diverse file formats. For example, the \textit{ParquetReader} is responsible for reading and parsing Parquet files. These file reader classes access data cached in Presto local cache, where data is systematically organized in pages (1 MB by default). In cases where the data is already cached (warm cache), it is promptly returned to the reader. Otherwise, the local cache employs a read-through strategy, loading the data from external storage for the reader. Operational insights reveal that parsing complex column-oriented data files can consume as much as 30\% of CPU resources \cite{wang2022metadata}. To mitigate the issue, Presto local cache also caches file metadata, including indexes, stripe (ORC) and column (Parquet) metadata, and file-level metadata. Thanks to its extensible design, the metadata, represented as key-value pairs, can be stored in memory, files, or persistent key-value stores like RocksDB. In contrast, data can be stored in memory or files, but not in external key-value stores due to its organization in data chunks (pages). In enterprise-grade production environments, data is usually cached in files and metadata in memory or RocksDB.

To ensure cache coherence, Presto will always fetch the latest metadata of input files from persistent storage, before splitting the input files on different workers along with the metadata. In case an input file is changed, the stale copy in the cache will be invalidated based on the timestamp of file creation or modification stored in the cache.

\subsubsection{Soft-Affinity Scheduling}

\begin{figure}[thp]
    \centerline{\includegraphics[width=0.4\textwidth]{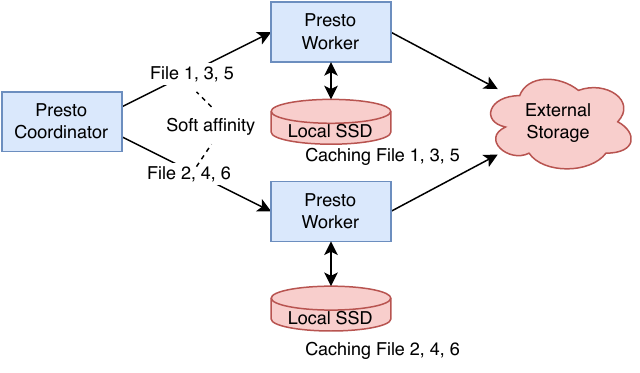}}
    \caption{File-level soft-affinity scheduling.}
    \label{fig.file-affinity}
\end{figure}

The Presto coordinator node is responsible for distributing splits to worker nodes, which in turn retrieve data from external storage for subsequent processing. Conventionally, each data file comprises multiple splits, and the scheduler's primary objective was to evenly distribute tasks by randomly assigning splits to workers. This approach, however, proved to be inefficient for caching as it led to frequent admission and eviction of data from each worker's local cache. To address that, we implemented a soft-affinity scheduler in Presto to allocate splits from the same file to the same worker node with the best efforts, as depicted in Figure \ref{fig.file-affinity}.

The soft-affinity scheduler uses the consistent hashing algorithm, with the file as the hashing input, to calculate the preferred worker node for a split. However, due to the potential for hot spots--where certain files are more frequently accessed than others--a designated worker node may be overwhelmed with an excessive number of splits. To gauge the workload of a node, the scheduler compares \textit{max-splits-per-node} with \textit{max-pending-splits-per-task}. If the initially chosen worker node is deemed busy, the scheduler opts for a secondary worker node from the hash ring. If the secondary node also lacks sufficient resources, it indicates a temporary inability to maintain soft-affinity, compelling the scheduler to assign the task to the least burdened worker in the cluster. This worker is instructed to fetch data directly from external storage, bypassing local caching, thus serving as a fallback.

\begin{figure*}[t]
    \centerline{\includegraphics[width=0.85\textwidth]{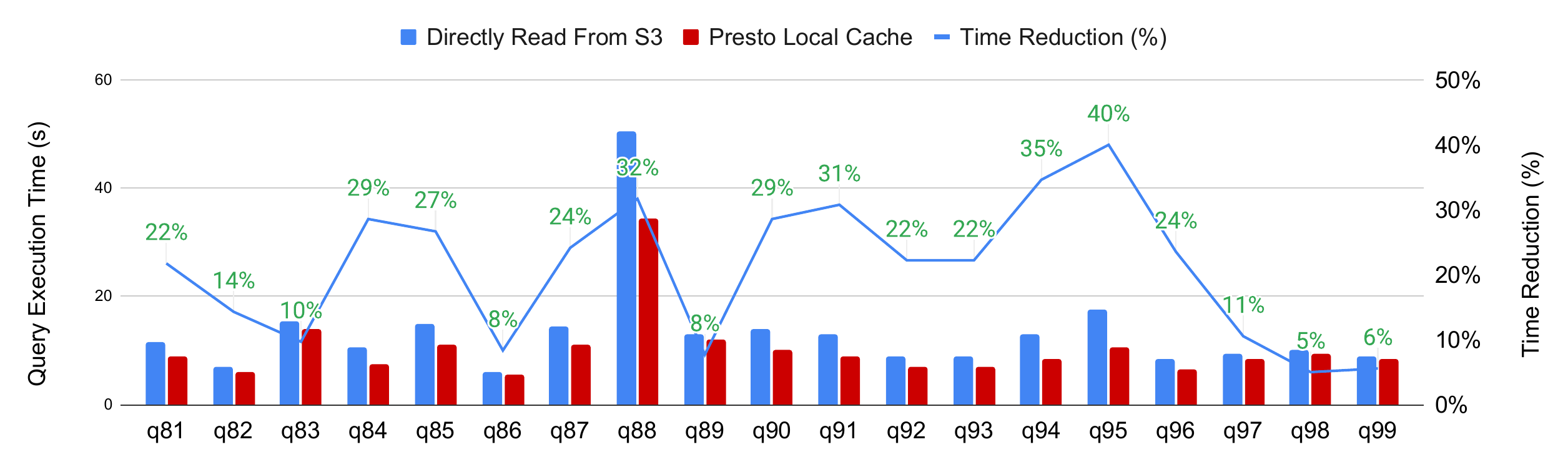}}
    \caption{Comparison of query execution time of TPC-DS Query 81 to Query 99 without and with Presto local cache.}
    \label{fig.presto-eval-tpcds}
\end{figure*}
\subsubsection{Per-Query Metrics Aggregation}

Our operational experience of Presto local cache underscores the importance of collecting and monitoring the metrics associated with frequently accessed data. In an enterprise-grade OLAP workload, access to data within a commonly used table is often unevenly distributed across its partitions, with some partitions experiencing substantially higher access frequencies. The local cache addresses this by identifying such hot partitions and aggregating per-query metrics to generate table-level insights. To facilitate this, we have developed a comprehensive metric system that integrates both Presto and Alluxio local cache metrics. Apart from utilizing Presto JMX exporters for logging JMX metrics, we have enhanced Presto \textit{RuntimeStats} to reflect the runtime statistics of each query stage.
Whenever Presto I/O operations engage the local cache, relevant metrics, such as cache hit rate and pages read, are recorded through the Alluxio system metrics. Concurrently, query-level runtime statistics are logged as in-memory metrics, which are periodically gathered for extensive monitoring and analytical purposes.

\subsubsection{Evaluation}

\textbf{TPC-DS benchmark.} To empirically validate the effectiveness of Presto local cache, we deployed a Presto cluster within the AWS Kubernetes environment, consisting of one coordinator node and four worker nodes. Each node has 32 CPU cores, 128 GB memory, and 150 GB SSD storage. The evaluation utilized the TPC-DS \cite{presto-tpcds} benchmark with a scale factor of 100. The dataset, stored in Parquet format, was hosted on AWS S3. We compared the query execution time between scenarios using the Presto local cache (warm cache) and direct reads from S3 (non-cache read). Figure \ref{fig.presto-eval-tpcds} indicates a reduction in query execution times of Query 81 to Query 99, ranging from approximately 10\% to 30\% when data is pre-loaded into the cache. The full TPC-DS benchmark evaluation results are included in Appendix \ref{appendix.tpc-ds}.

\begin{figure}[htb]
    \centering
    \begin{subfigure}[t]{0.23\textwidth}
        \includegraphics[width=\linewidth]{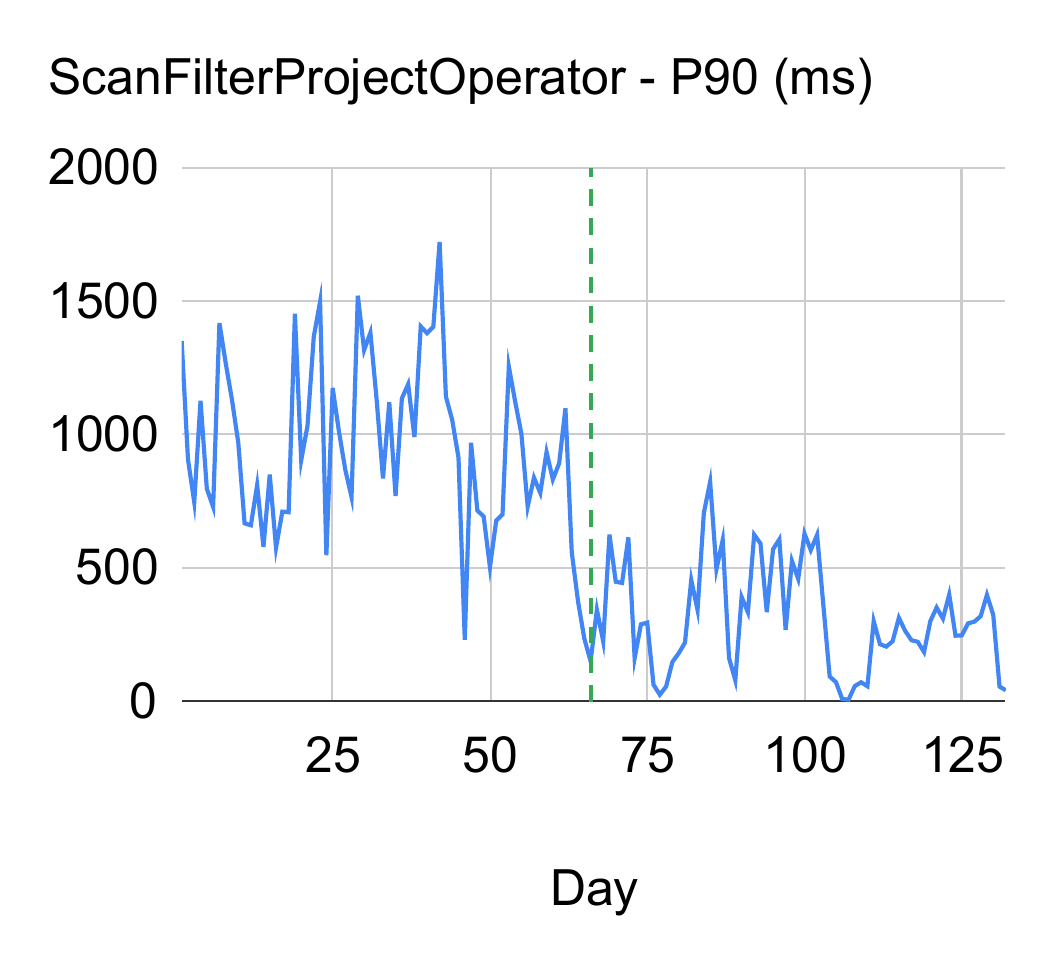}
        \caption{90th percentile (P90) of query time spent on reading files before and after enabling the cache.}
        \label{fig.presto-inputwall90}
    \end{subfigure}
    \hfill
    \begin{subfigure}[t]{0.23\textwidth}
        \includegraphics[width=\linewidth]{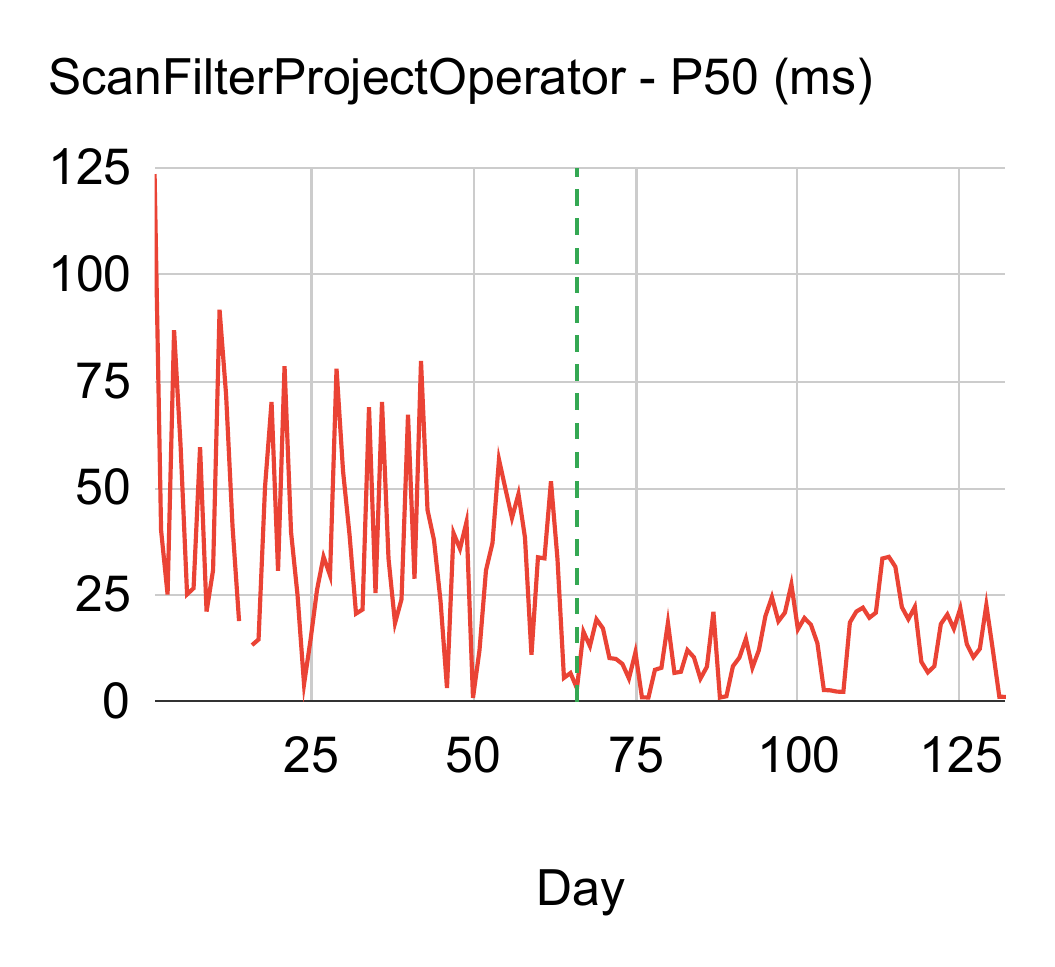}
        \caption{50th percentile (P50) of query time spent on reading files before and after enabling the cache.}
        \label{fig.presto-inputwall50}
   \end{subfigure}
   \caption{Comparison of execution time spent on fetching data before and after enabling the cache.}
\label{fig.presto-inputwall}
\end{figure}

\textbf{Uber's production results.}
In an evaluation within Uber's enterprise-grade production environment, we focused on the performance enhancement in query processing attributable to the utilization of local caching. This assessment involved measuring the time spent on reading Parquet files and comparing the results obtained before and after the activation of caching. In Presto, this is quantified by the \textit{inputWall} metric of the \textit{ScanFilterProjectOperator}, a key internal phase within a Presto query, responsible for data input handling and initial filtering. Notably, at Uber, for the tables onboarded with Presto local cache, more than 97\% of queries involve reading through a ScanFilterProjectOperator stage, underscoring the relevance of this metric. Figure \ref{fig.presto-inputwall} shows that the 90th percentile (P90) of the time spent on reading files was reduced by 67\% and 50th percentile (P50) was decreased by 64\%.

\textbf{Meta's production results.} Similar performance enhancements were observed in Meta's production data analytics. Specifically, in one of Meta's internal use cases, the query latency P50 was reduced by around 33\%, and P95 was reduced by around 49\% \cite{meta-alluxio}. These results highlight Presto local cache's effectiveness in mitigating the query processing tail latency. Additionally, there was a 57\% reduction in total data scanned from remote storage, indicating the local cache's impact in reducing data transfer between compute and storage layers.

\subsection{Case Study - HDFS Local Cache}\label{sec.use-case.hdfs}

\subsubsection{Design Overview}

\begin{figure}[htb]
    \centerline{\includegraphics[width=0.35\textwidth]{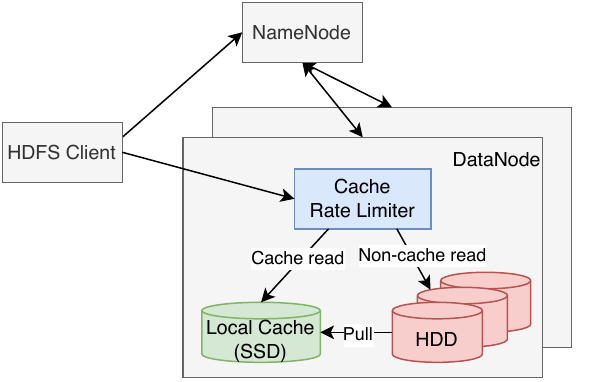}}
    \caption{The architectural design of HDFS local cache.}
    \label{fig.hdfs-local-cache}
\end{figure}

The Alluxio local cache module is embedded into each HDFS DataNode to construct the HDFS local cache, as depicted in Figure \ref{fig.hdfs-local-cache}. The local cache stores hot data in the local SSD and original datasets are stored in the HDD due to the capacity restrictions. Take Uber's production HDFS clusters as an example, with 500 TB of data storage capacity on a DataNode, only a few TBs of SSD is available \cite{uber-hdfs}. It is unfeasible to cache all data, which will cause frequent cache admission and eviction and lower cache hit rates. To ensure cache efficiency, when a requested data block is not loaded in the cache, the cache rate limiter will decide whether the data block should be loaded into the cache, based on historical data access patterns. If the data block is unsuitable for caching, the cache rate limiter will go to the normal non-cache read path. Otherwise, if the data block is determined as cache-worthy, this block will be cached for faster access in the future.

Similar to the implementation of Presto local cache, HDFS local cache employs the Hadoop-compatible file system APIs provided by Alluxio, which help with the seamless integration with the DataNode. Notably, the soft-affinity scheduling in Presto local cache does not apply here as the HDFS NameNode has already maintained a metadata table, recording the location of each data block.

HDFS DataNodes manage HDFS blocks, each contains a block file and a metadata file. The metadata file contains checksum info of the block file, and therefore, it has to match the corresponding block file all the time. Upon caching a block on disk, both the block and metadata files are stored separately on the local SSD, maintaining the same format as the original files. This design guarantees that either both the block and metadata files are read from the cache, or both are read from their original non-cache locations, but never any form of the mix. This block-level caching design simplifies reasoning and enhances reliability, forming the basis for subsequent design decisions discussed below.

\subsubsection{Cache Rate Limiter}

The cache rate limiter plays a crucial role in favoring the retention of frequently accessed data blocks in the cache while restricting those less frequently accessed. By evaluating a few design options against simulation tests based on production traffic, we created the rate limiter with the design we called \textit{BucketTimeRateLimit}.

\begin{figure}[htb]
    \centerline{\includegraphics[width=0.47\textwidth]{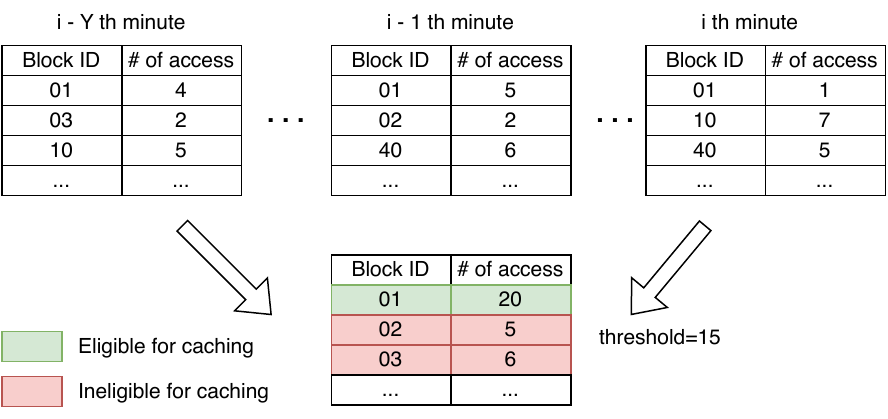}}
    \caption{Data structures used in the cache rate limiter.}
    \label{fig.hdfs-local-cache-window}
\end{figure}

The algorithm functions by evaluating if a data block has been accessed more than X times in the past Y time interval. It incorporates an ordered list of buckets as a sliding window. Each bucket logs the access count for each block during its respective minute-long window. BucketTimeRateLimit keeps a constant number of active buckets and discards the oldest bucket every minute. To determine whether a block should be subjected to rate limiting, the algorithm aggregates the total number of accesses across all existing buckets, as shown in Figure \ref{fig.hdfs-local-cache-window}. A block is classified as high-frequency and eligible for caching if its access count surpasses a predefined threshold (e.g. 15 in the example).

\subsubsection{HDFS Write Operations}

Although the HDFS local cache stores only finalized block files, certain HDFS operations can still modify these blocks, necessitating updates in the cached data. Such HDFS operations include appending to a block (\textit{append}), deleting a block (\textit{delete}), and truncating a block (\textit{truncate}). Since HDFS \textit{truncate} is not used in our HDFS local cache use cases, we focus on the \textit{append} and \textit{delete} operations.

\textbf{Append to a block.} Within HDFS, the append operation allows a client to add new content to the end of an existing file. To distinguish between the original and modified blocks, HDFS employs a versioning system where each block is assigned a \textit{generation stamp}. Each invocation of the append operation increments the block's generation stamp, signaling a new version of the block.

HDFS local cache uses the generation stamp for snapshot isolation. When a block and its checksum are loaded into the cache, the associated generation stamp is retained and used in conjunction with the blockID to create a unique cache key. This approach ensures that during an ongoing append operation, cache readers access the original version of the block, as our design prohibits loading blocks that are currently being written into the cache. Post completion of the append operation, the updated block, identifiable by its new generation stamp, is considered a distinct cache entry. This allows the latest version of the block to be cached and made available for subsequent read operations.

\textbf{Delete a block.} The deletion of a block in HDFS requires the removal of its cache copy. Initially, our approach involved a time-based policy for periodically purging obsolete cache entries. To improve the efficiency and delete outdated cache entries more timely, we introduced an in-memory mapping within each DataNode. This mapping tracks the blocks currently cached, enabling the DataNode to identify and instruct the local cache to eliminate related page files upon block deletion. The in-memory mapping takes the form of \textit{<blockId -> (cacheId, fileLength)>}, where \textit{fileLength} helps compute the relevant page files.

A caveat of this in-memory approach is the loss of mapping data when a DataNode restarts. To address this, the DataNode clears all local cached contents and rebuilds the cache from the ground up upon restarting. Although this approach may impact cache efficiency, we deem it a viable compromise, given that DataNode restarts are infrequent in production.

\subsubsection{Evaluation}

\begin{figure}[htb]
    \centerline{\includegraphics[width=0.4\textwidth]{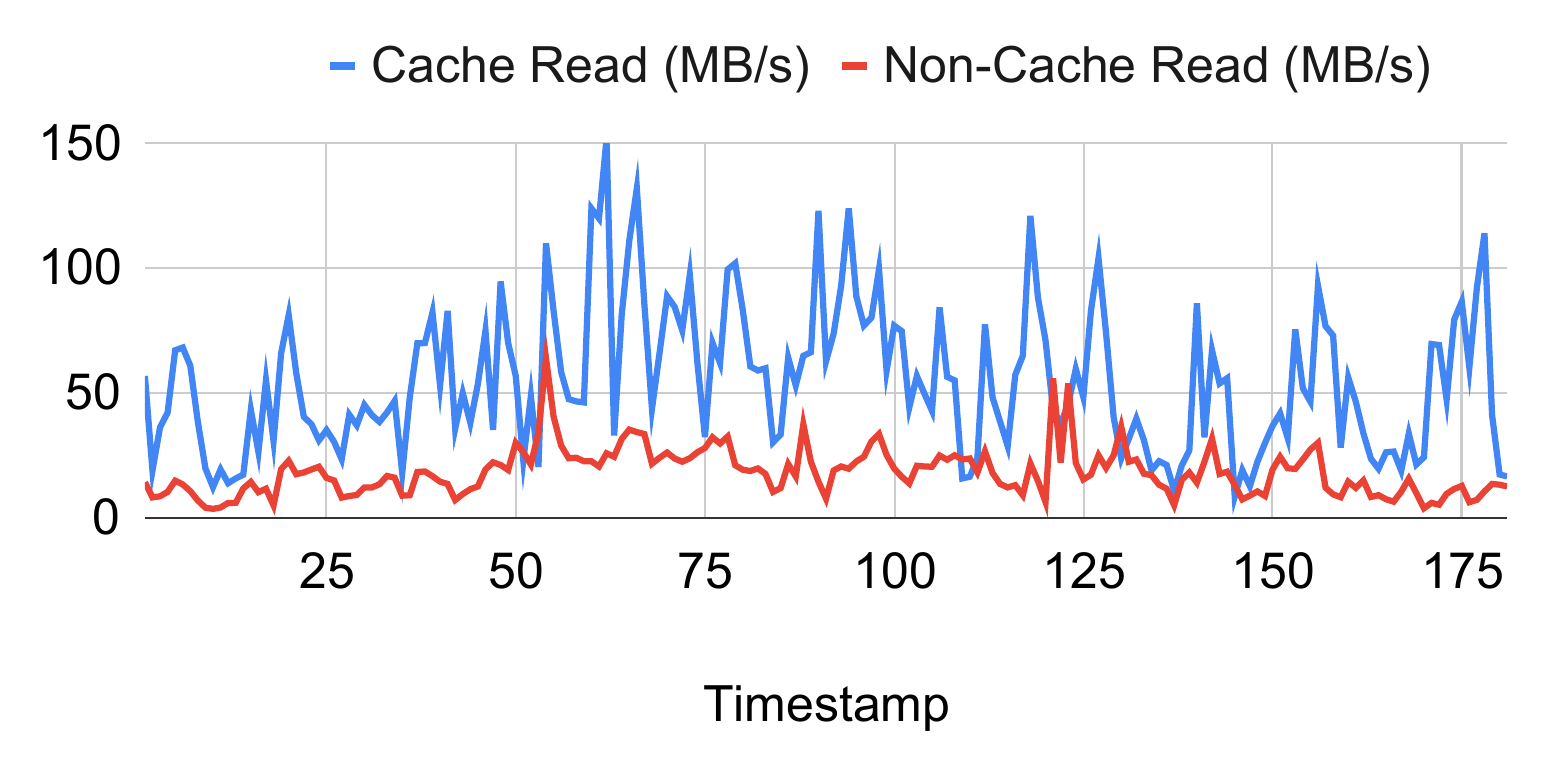}}
    \caption{The comparison of cache read rates and non-cache read rates in an HDFS DataNode.}
    \label{fig.hdfs-cache-read}
\end{figure}

To validate the effectiveness of HDFS local cache, we deployed it within Uber's production HDFS clusters. As depicted in Figure \ref{fig.hdfs-cache-read}, the introduction of HDFS local cache led to a substantial portion of read traffic being handled by the cache. This observation aligns with our earlier workload analysis in Section \ref{sec.background.workload}, which identified that a significant share of traffic is directed towards hot spots. In this DataNode, in the given time period of one hour, the rate of bytes read from the cache is, on average, threefold that of non-cache reads. More than 70\% of total read bytes are serviced by the local cache.

\begin{figure}[htb]
    \centerline{\includegraphics[width=0.36\textwidth]{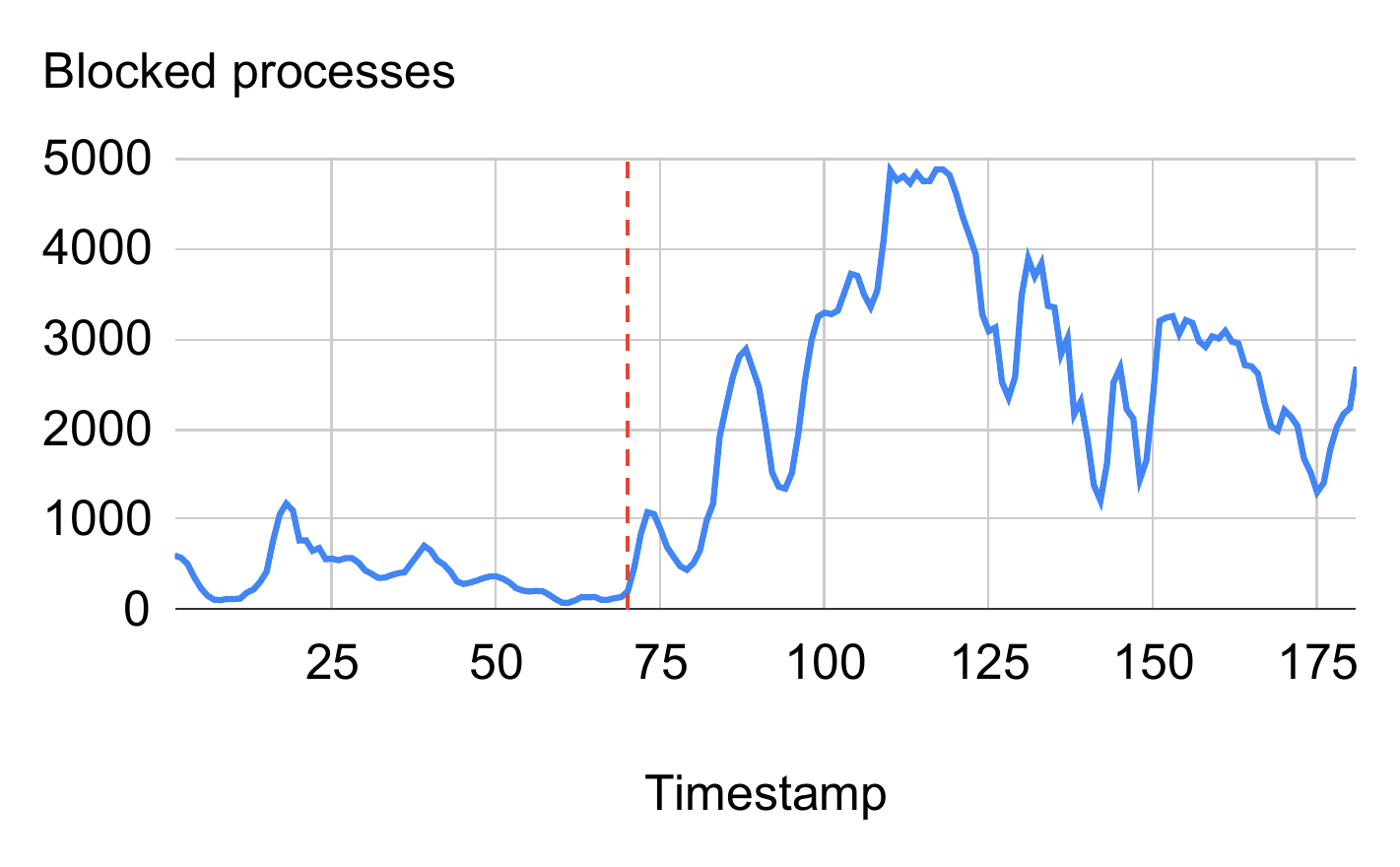}}
    \caption{The comparison of blocked processes (I/O throttling) in one DataNode enabling and disabling the local cache.}
    \label{fig.hdfs-blocked-process}
\end{figure}

Section \ref{sec.background.workload} highlights the I/O throttling issue in the production HDFS cluster, reflected in a high number of blocked processes in the DataNodes. Figure \ref{fig.hdfs-blocked-process} compares the number of blocked processes in a DataNode, with and without the HDFS local cache. Upon disabling the cache at timestamp 70, there is a rapid increase in blocked processes, reaching up to approximately five thousand. During this one-hour period, the local cache reduces the number of blocked processes by an average of 86\%.

\section{Lessons Learned}\label{sec.lessons}

This section recounts the experiences and qualitative lessons gleaned from the development and operation of Alluxio local cache in organizations like Alluxio, Uber, and Meta.

\textbf{Enhancing OLAP system performance with data file metadata caching.}
Nowadays, column-oriented file formats, such as Apache Parquet and ORC, are widely used in enterprise-grade data analytical systems. These formats involve extensive scanning and parsing of metadata by query systems, consuming up to 30\% of CPU resources in practice, thereby impacting overall system performance. Our experience shows that caching metadata, like column metadata in Parquet files, can significantly alleviate this overhead. Specifically, caching deserialized metadata objects can reduce CPU usage by up to 40\%, enhancing the efficiency of query systems \cite{wang2022metadata}.

\textbf{The trade-off in cache page size between read amplification and the number of requests to remote storage.}
Our analysis, as detailed in Section \ref{sec.background.workload}, indicates that large-scale query systems like Presto are inclined to generate a large number of small read operations. A larger cache page size, while reducing the number of read requests to remote storage, increases read amplification. Conversely, smaller cache page sizes reduce data fetched but increase the metadata memory footprint and the number of storage requests.

In practice, either a high read amplification or a large number of requests to storage can have a non-negligible impact on the organizational-level infrastructure resource utilization and system performance, potentially leading to unexpectedly high monetary costs. The ideal page size is a function of traffic patterns, file formats, and desired metadata size. Our empirical findings, particularly in the context of Uber and Meta's production traffic, indicated that a cache page size of 1 MB strikes an optimal balance. For practical cache page size tuning, we recommend starting with a page size of 1 MB or a number akin to the data file's page size, followed by gradual adjustments to identify the optimal setting.

\textbf{An aggregated metrics system is crucial for cache tuning and debugging.}
A modern enterprise-grade data analytical or storage system can contain thousands of nodes, leading to thousands of local cache deployments. An important design decision we made is implementing an aggregated metrics system, collecting both basic metrics from the local cache and aggregated metrics based on the applications. The metric system provides a centralized view of predefined and user-customized metrics and helps developers drill down to detailed metrics. We found such a system is helpful for cache setting tuning and debugging. Besides the common metrics, such as cache hit rate, cache eviction rate, and cache bytes read from storage, we also learned that error-related metrics, including error counts of different operations (put, get, delete, etc.) and breakdowns of concrete types of errors, are extremely helpful to identify root causes in debugging.

\textbf{Keeping the seats for temporary offline nodes (lazy data movement).}
Nowadays, enterprise-grade large-scale systems are usually deployed in container orchestration platforms, such as Kubernetes \cite{verma2015large} and Mesos \cite{hindman2011mesos}. Handling temporary offline nodes becomes a critical aspect of cache architecture design. These systems, due to various operational factors such as application failures, resource constraints, periodic maintenance, or software updates, may experience intermittent node restarts. To effectively manage such scenarios, we incorporated a strategy in our cache design that utilizes a timeout mechanism within the hashing ring. This approach is designed to allocate a waiting period for nodes that temporarily go offline. If these nodes come back online within the timeout period, the system avoids unnecessary data movement across the cache nodes. This method of 'lazy data movement' has proved to be efficient in maintaining cache stability and performance, mitigating the impact of frequent data movement in containerized deployment environments.

\textbf{Considering data imbalance and fallback when setting the number of cache replicas.}
Hot spots are common in production data traffic. Some tables or data blocks are much more frequently accessed than others. This uneven distribution of data access stimulates careful consideration in the allocation of cache replicas. Our operational insights suggest that increasing the number of replicas can alleviate pressure on hot spots but may inadvertently lead to increased latency in locating an unoccupied cache node. In practice, to find an optimal balance, we adopted a strategy that limits the number of cache replicas to a maximum of two. In cases where both replicas are unavailable due to resource limitations, the system defaults to retrieving data from remote storage. This hybrid approach, which combines a limited replica count with a remote storage fallback, has demonstrated greater robustness and lower latency in practice compared to simply increasing the number of replicas. This method effectively addresses the challenges posed by data hot spots while optimizing for system performance and reliability.

\textbf{Trade-offs between local cache and distributed cache.}
At Meta and Uber, implementing a local cache was favored given the stringent timeline while adding services in production is more challenging. Moreover, local caching offers a substantial return on investment (ROI) for decision-makers by efficiently utilizing available resources on compute nodes. For example, each of Meta's Presto node is capable of utilizing 1-2 TB of SSD resources for local caching. However, local caching also has certain limitations. It is constrained by the available local resources, lacks the ability to scale independently, and does not support data sharing across different computing engines, such as between Presto and Spark.

\section{Failure Case Study}\label{sec.failures}

This section shares insights from unexpected failures encountered in the operation of the local cache within large-scale distributed systems, such as Presto and HDFS.

\textbf{File read hanging.}
As we discussed in Section \ref{sec.arch.storage}, Alluxio local cache employs the local SSDs to store the cached data. Accessing the cached data requires a \textit{read\_file} operation from the disk. From our operational experience, the local cache sometimes experiences read operation hangups, for as much as 10 minutes, probably caused by resource contention with other applications on the shared machine. Implementing a fallback to the remote storage system and with a default 10-second timeout for \textit{read\_file} operations have proven to be effective in mitigating this issue. If the operation does not return in this time range, the local cache will retrieve the data from the remote storage.

\textbf{Corrupted files.}
In the realm of SSD-based caching systems, such as the one used in Presto local cache, file corruption, although infrequent, remains a potential issue. Instances of corrupted files have been observed during our operations, leading to significant impacts such as the failure of SQL queries accessing these files and a notable decrease in cache hit ratios due to the cache's inability to write new data. To address this challenge, we adopted a proactive strategy by initiating early eviction of cache entries when the failure occurs, even before the cache reaches its full capacity. This approach has been instrumental in mitigating the adverse effects of file corruption. Furthermore, as highlighted in the previous section, a detailed breakdown of error types played a crucial role in identifying and understanding the root causes of such failures.

\textbf{Insufficient disk capacity.}
Alluxio local cache has a configurable cache capacity that is determined during the cache start. This configuration defines the maximum disk space allocated for cache use. However, it is possible for the SSD storage to become fully utilized before reaching the set cache capacity during ongoing operations. Such scenarios of insufficient disk capacity can lead to intricate and demanding debugging processes. To address the issue, through catching the \textit{No space left on device} exception, the local cache initiates early cache eviction, rather than waiting for the cache to reach its pre-configured full capacity.

\section{Related Work}\label{sec.related-work}

The growing need for an efficient cache system for large-scale data analytics and storage has stimulated a large body of research work. We highlight the most closely related work.

\textbf{Cache for OLAP.} Caching saves data access from remote storage in OLAP systems by keeping the frequently accessed data in memory or high-bandwidth fast storage. NoDB \cite{alagiannis2012nodb} recognizes raw data files as first-class citizens and uses caching to optimize query execution on external tables. Many also find caching helpful for bridging the performance gap caused by storage-compute disaggregation architecture in modern cloud databases. FlexPushdownDB by Yang et al. \cite{yang2021flexpushdowndb} aims to keep hot data in the compute layer and reduce the amount of data transferred between storage and compute. Similarly, Crystal by Durner et al. \cite{durner2021crystal} provides a unified cache management system as a storage middleware to improve query latency and save bandwidth costs between compute and storage in cloud analytical databases. Apache Spark initially used Tachyon \cite{li2014tachyon} (now rebranded as Alluxio) for built-in resilient distributed datasets (RDD) caching. Alluxio local cache was developed by the original Tachyon team, with a special focus on a simple, independent, and embedded caching system. A similar caching framework is Rubix \cite{rubix}, which provides a lightweight data cache for big data engines. Compared with Rubix, Alluxio local cache operates without the need for daemon processes (\textit{BookKeeper}) and employs soft-affinity scheduling for higher cache hit rate and performance.

\textbf{Cache for distributed file systems.} Caching is also crucial for distributed file systems to reduce latency by keeping hot data blocks or files in faster storage. HDFS implements a centralized cache management in which the NameNode instructs DateNodes to cache hot data blocks in off-heap memory \cite{hdfs-cache}. This HDFS caching contributes to the construction of multi-tier storage in distributed file systems. For example, Herodotou et al. \cite{herodotou2019automating} presented a distributed tier-based storage system using HDFS cache as one of the tiers. Furthermore, as SSDs outperform conventional HDDs, similar to the HDFS local cache presented in this paper, Zhao et al. \cite{zhao2013hycache} created a unified caching middleware for distributed file systems by organizing the storage in both HDDs and SSDs.

\section{Conclusion}\label{sec.conclusion}
Designed to tackle a range of challenges, including massive-scale data-intensive analytics, uneven data access distribution, limited I/O bandwidth, and fragmented access in column-oriented files, the relatively straightforward implementation of Alluxio local cache has proven its notable effectiveness in enterprise-grade petabyte-scale OLAP and storage systems. Its application as the Presto local cache has led to a 49\% reduction of query latency P95 and a 57\% decrease in data retrieval from remote storage within Meta's production data analytics. HDFS local cache, adopted in Uber's production HDFS clusters, resulted in an 86\% reduction in blocked processes due to I/O throttling. The insights and lessons gathered from the development and operation of these cache systems are believed to be of substantial value for building enterprise-grade compute and storage systems.

\section*{Acknowledgment}

The widespread adoption of Alluxio local cache highlights a successful collaborative effort involving numerous individuals. We would like to express our gratitude to Haoyuan Li, Amelia Wong, Xiaodan (Danica) Wang, Xuan Du, Adit Madan, Rongrong Zhong, Rohit Jain, Junyan Guo, Fengnan Li, and many others. We also thank the anonymous USENIX ATC reviewers for their insightful comments, which considerably improved this paper.

\bibliographystyle{plain}
\bibliography{library}

\clearpage
\appendix

\section{TPC-DS Evaluation}\label{appendix.tpc-ds}

\vspace{-0.4em}
\begin{figure}[H]
    \centering
    \includegraphics[width=1.1\textwidth,angle=-90]{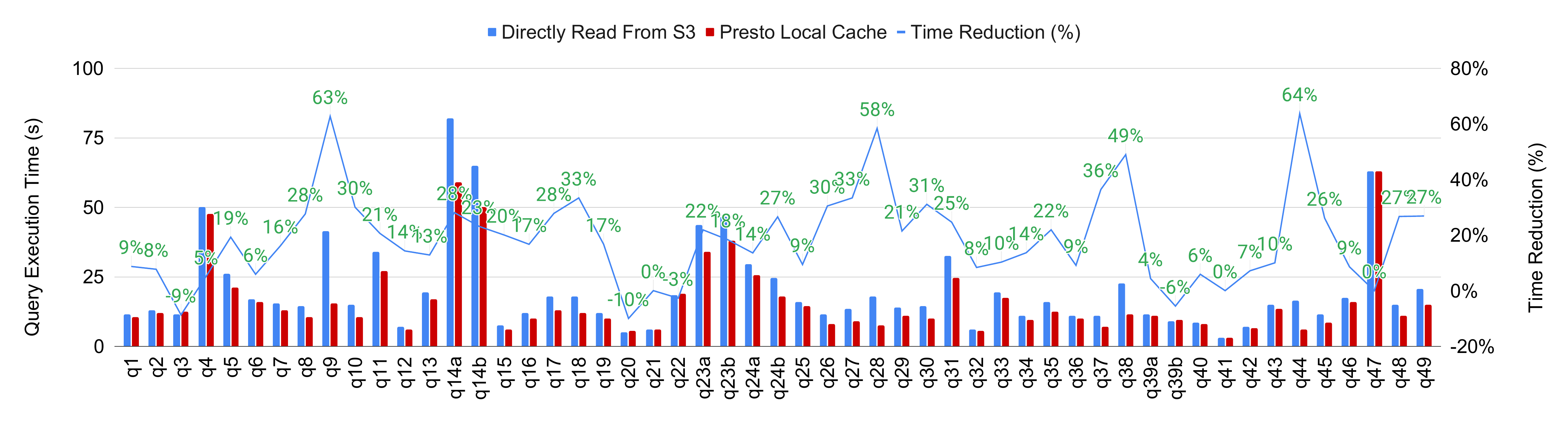}
    \caption{Comparison of query execution time of TPC-DS Query 1 to Query 49 without and with Presto local cache.}
    \label{fig.tpc-ds-full1}
\end{figure}

\begin{figure}[h]
    \centerline{\includegraphics[width=1.1\textwidth,angle=-90]{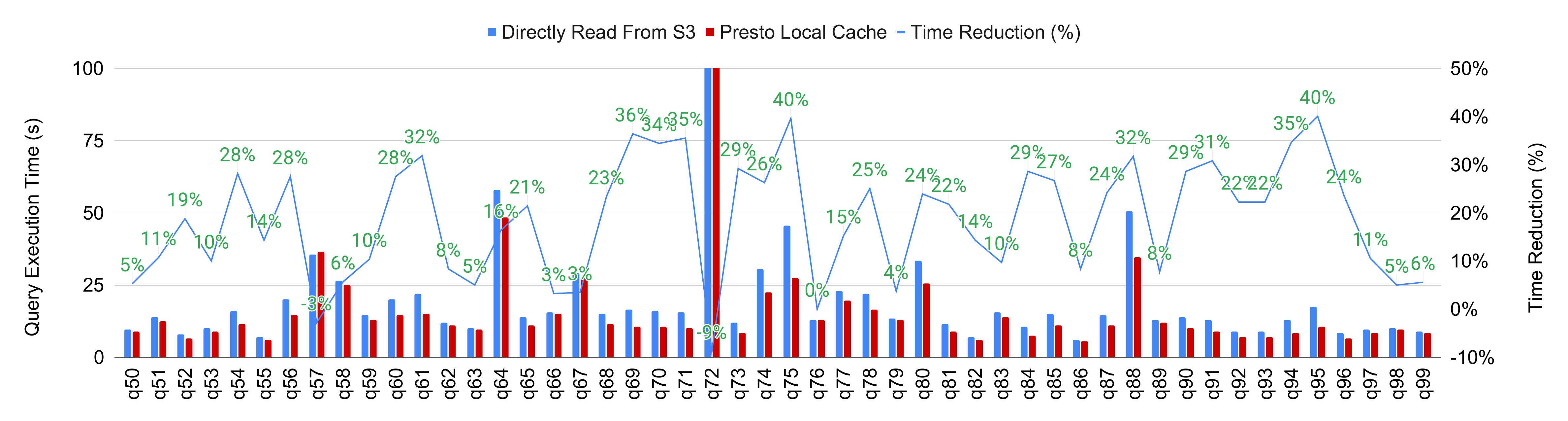}}
    \caption{Comparison of query execution time of TPC-DS Query 50 to Query 99 without and with Presto local cache.}
    \label{fig.tpc-ds-full2}
\end{figure}

\end{document}